\documentclass[prd,onecolumn,noshowkeys,noshowpacs,nofootinbib,12pt]{revtex4-1}
\usepackage{amsmath}
\usepackage{amssymb}
\usepackage{amsfonts}
\usepackage{amsthm}
\usepackage{mathtools}
\usepackage{bm}
\usepackage{hyperref}

\usepackage{graphicx}
\usepackage[T1]{fontenc}

\theoremstyle{plain}

\theoremstyle{definition}

\theoremstyle{remark}

\begin{document}

\title{An Optical Geometry Perspective on \\ 
the Emerging Conformal Field Theory \\ Behavior on the Cosmological Event Horizon: \\ 
Massless Fermions on de Sitter Space}

\author{Pawel  O. Mazur, Pawel J. Morawiec}
\affiliation{Department of Physics and Astronomy, University of South Carolina, Columbia, SC 29208}
\begin{abstract}
\noindent We have investigated the emerging conformal field theory behavior on the interior interface 
of the region near the cosmological event horizon on an example of massless Dirac fermions. 
Indeed, we have constructed and verified the emerging de Sitter/CFT correspondence 
for the near cosmological event horizon region. 
\par\noindent
It is in this near horizon region where the infinitely thin shell of matter with the stiffest equation of state (Zeldovich's matter) is placed in the model for a thermodynamically stable 
`would be black hole' which is sometimes called a gravastar \cite{MaMotPNAS} or a dark energy star. 
We put forward the hypothesis to the effect that the Zeldovich matter consists 
of the strongly correlated quanta which are best described by the emerging conformal field theory 
in the near `would be horizon' region of a gravastar/dark energy star \cite{MaMotPNAS}. 
 
\end{abstract}
\date{June 2010}
\maketitle

\section{Introductory remarks}\label{sec:OGIntro}
\noindent
It has been known for many years that the special relationship between quantum field theories on $(d+1)$-dimensional de Sitter (and anti-de Sitter) space and $d$-dimensional conformal field theories exists \cite{TagirovChernikov65, Staruszkiewicz:1987wt,Antoniadis:1997fu,MaMot_PR_D2001,Casher:2003gc,Antoniadis:2006wq,Witten:1998qj}. 
In order to see this relationship consider for illustrative purposes a free massive scalar field $\Phi$ on the $(d+1)$-dimensional de Sitter space with the metric given in spatially flat coordinates
\begin{equation}
ds^{2} = d\tau^{2} - e^{2H\tau}d\vec{x}^{2}
\end{equation}
which is described by the action
\begin{equation}
I = \frac{1}{2}\int d\tau (dx)_{d}\,\sqrt{-g}\left(g^{\mu\nu}\partial_{\mu}\Phi \partial_{\nu}\Phi  - m^{2}\Phi^{2}\right)
\end{equation}
where $(dx)_{d}$ is the volume element on the flat Euclidean space $\mathbb{R}^{d}$.

Quantizing this massive scalar field one can compute the Wightman two-point correlation function
\begin{equation}
G(x, x^{\prime}) = <\Phi(x) \Phi(x^{\prime}) >\qquad .
\end{equation}

It has been shown \cite{TagirovChernikov65, Staruszkiewicz:1987wt,Antoniadis:1997fu,MaMot_PR_D2001,Casher:2003gc,Antoniadis:2006wq,Witten:1998qj} that the asymptotic behavior of $G(x, x^{\prime})$ for $\tau, \tau^{\prime}\rightarrow\infty$  is \cite{Antoniadis:1997fu}

\begin{eqnarray}
&&<\Phi(x)\Phi(x')>\sim A\,{D(x,x')}^{-2\Delta_{+}} +B\,{D(x,x')}^{-2\Delta_{-}} \label{twopt}\\
&&\Delta_{\pm} = \frac{d}{2} \pm \sqrt{\frac{d^{2}}{4} - {m^2\, R^2}} \label{Delta}
\end{eqnarray}
where $R=H^{-1}$ is the radius of curvature 
  and  $D(x,\, x')$ is the geodesic distance between two points $x$ and $x'$. In spatially flat coordinates

\begin{equation}
D^2(x,x') = e^{H(\tau+\tau')}
\left[H^{-2}(e^{-H\tau} - e^{-H\tau'})^2 - ({\bf x}
- {\bf x'})^2\right].
\label{sigma}\end{equation}

For equal times $\tau=\tau'$ the asymptotic behavior of the two-point correlation function \eqref{twopt} is the following
\begin{equation}
\label{CFT}
G({\bf x}, {\bf x}') \sim A' |{\bf x}-{\bf x'}|^{-2\Delta_{+}} + B' |{\bf x}-{\bf x'}|^{-2\Delta_{+}}
\end{equation}

The power law behavior in \eqref{CFT} is the characteristic property of  the Euclidean Conformal Field Theory (CFT) in $d$-dimensions.

In Euclidean Conformal Field Theory the two-point correlation function of two primary operators 
$\mathcal{O}_{\Delta_{1}}$ and  $\mathcal{O}_{\Delta_{2}}$ and the three-point function of primary operators 
$\mathcal{O}_{\Delta_{1}}$, $\mathcal{O}_{\Delta_{2}}$ and $\mathcal{O}_{\Delta_{3}}$ 
is completely fixed by conformal invariance \cite{Polyakov:1970xd}
\begin{equation}
\label{cft1}
<\mathcal{O}_{\Delta_{1}}({\bf x_{1}})\; \mathcal{O}_{\Delta_{2}}({\bf x_{2}}) >_{CFT} = A(\Delta_{1})\delta_{\Delta_{1},\Delta_{2}} |{\bf x_{1}}-{\bf x_{2}}|^{-2\Delta_{1}}
\end{equation}

\begin{multline}
\label{cft2}
<\mathcal{O}_{\Delta_{1}}({\bf x_{1}})\; \mathcal{O}_{\Delta_{2}}({\bf x_{2}})\; \mathcal{O}_{\Delta_{3}}({\bf x_{3}}) >_{CFT} = B(\Delta_{1},\Delta_{2},\Delta_{3}) \times
\\ |{\bf x_{1}}-{\bf x_{2}}|^{-\Delta_{1}-\Delta_{2}+\Delta_{3}}\,|{\bf x_{2}}-{\bf x_{3}}|^{-\Delta_{2}-\Delta_{3}+\Delta_{1}}\,|{\bf x_{3}}-{\bf x_{1}}|^{-\Delta_{3}-\Delta_{1}+\Delta_{2}} \qquad\qquad
\end{multline}

The appearance of the CFT two-point function \eqref{cft1} in the asymptotic behavior of the two-point function for the massive scalar field \eqref{twopt} and \eqref{CFT} can be understood in terms of an isomorphisms of the invariance groups. In Quantum Field Theory (QFT) on de Sitter space the two-point function is an invariant function with respect to the isometry group of de Sitter space. In Euclidean CFT on $d$-dimensional Euclidean space the two-point function is invariant with respect to the conformal group of Euclidean space $C(d)$.

The conformal group $C(d)$ of the Euclidean space in $d$-dimensions is $SO(1,\,d+1)$ which is at the same time the isometry group of $(d+1)$-dimensional de Sitter space. The group $SO(1,\,d+1)$ acts transitively on its orbits in $(d+2)$-dimensional Minkowski space. There are three distinct orbits of $SO(1,\,d+1)$ in the Minkowski space: the de Sitter spacetime, the light cone and the Lobachevski space. The future infinity on the de Sitter hyperboloid is asymptotic to the light cone from the outside, while the spatial infinity of the Lobachevski's hyperboloid is asymptotic to the light cone from the inside. The space of directions on the light cone is the projective space which is the $d$-dimensional sphere $S^{d}$. The Lorentz group $SO(1,\,d+1)$ acts on this space of directions $S^{d}$ as the conformal group $C(d)$. It is because the de Sitter hyperboloid and the Lobachevski hyperboloid asymptote to the sphere at infinity on the light cone on which the Lorentz group acts as the conformal group $C(d)$ that the natural action of the conformal group on spatial infinities of de Sitter and Lobachevski spaces emerges. It then must follow that the asymptotic behavior of two-point Green's functions on the spatial infinities of de Sitter and Lobachevski spaces be given by the conformally invariant     two point functions on those boundaries. This general statement has been illustrated on the example of the massive scalar field on de Sitter space where it was first discovered \cite{TagirovChernikov65, Staruszkiewicz:1987wt,Antoniadis:1997fu,MaMot_PR_D2001,Casher:2003gc,Antoniadis:2006wq,Witten:1998qj}.

The case of  a massless scalar field is rather special and the connection between Quantum Field Theory (QFT) on the 3-dimensional de Sitter space and the Euclidean CFT on its future time-like infinity $S^{2}$ was first discovered in 1987 by A. Staruszkiewicz \cite{Staruszkiewicz:1987wt}. In this case one considers the primary operators $e^{\imath \Phi(x)}$ for which the two-point function asymptotes to 
\begin{equation}
< e^{\imath \Phi(x)} e^{-\imath \Phi(y)}> \sim |{\bf x} -{\bf y}|^{-2\Delta}
\end{equation}
when $\tau = \tau' \rightarrow \infty$, with ${\bf x}, {\bf y}$ the stereographic coordinates on $S^{2}$.  Here $\Delta = e^{2}/\pi$ and $e$ is the dimensionless constant which appears in the action for the massless scalar field \cite{Staruszkiewicz:1987wt}
\begin{equation}
I = \frac{1}{8\pi\,e^{2}}\int d^{3}x\,\sqrt{-g}g^{\mu\nu}\partial_{\mu}\Phi\partial_{\nu}\Phi .
\end{equation}

\section{On the AdS/CFT connection}\label{sec:FermionsA}
We have already seen the emerging relationship between QFT on $(d+1)$-dimensional de Sitter space and the CFT on the $d$-dimensional sphere $S^{d}$ at $\tau \rightarrow +\infty$ on the examples of massive and massless spin-0 scalar fields  \cite{TagirovChernikov65, Staruszkiewicz:1987wt,Antoniadis:1997fu,MaMot_PR_D2001,Casher:2003gc,Antoniadis:2006wq}. The same relationship also emerges for vector and spinor fields. 
We have argued that the same relationship  between Euclidean QFT on $d+1$-dimensional Lobachevski space and Euclidean $d$-dimensional CFT on its spatial infinity $S^{d}$ must emerge on the basis of the geometrical picture and group isomorphism described above \cite{Staruszkiewicz:1987wt,MaMot_PR_D2001,Casher:2003gc}.

Indeed for a massive scalar field $\Phi$ on the $(d+1)$-dimensional Lobachevski space $\mathcal{M}$ with the metric
 \begin{equation}
ds^{2} = R^{2} \frac{dz^{2}+{\bf dx\cdot dx}}{z^{2}}
\end{equation}
where ${\bf x}$ is the coordinate on the $d$-dimensional Euclidean space and $R$ is the radius of curvature, the action is \cite{Witten:1998qj, Gubser:1998bc}
\begin{multline}
\label{actI1}
I = \frac{1}{2}\int_{\mathcal{M}} dz (dx)_{d}\,\sqrt{g}\left(g^{\mu\nu}\partial_{\mu}\Phi \partial_{\nu}\Phi  + m^{2}\Phi^{2}\right) = \\
\frac{1}{2}\int_{\mathcal{M}} dz (dx)_{d}\,\,\sqrt{g}\,\Phi (-D^{\mu}D_{\mu} + m^{2})\Phi + \frac{1}{2}\int_{\partial \mathcal{M}} (dx)_{d}\,\,\sqrt{h}\,n^{\mu}\Phi\partial_{\mu}\Phi\qquad\qquad
\end{multline}
where the boundary $\partial \mathcal{M}$ is at $z = \epsilon \rightarrow 0^{+}$, and $n^{\mu} = \frac{z}{R}\delta^{\mu}_{z}$, $\sqrt{h} = (\frac{R}{z})^{d}$. For solutions of the equation of motion 
\begin{equation}
\label{phiEq}
(-D^{\mu}D_{\mu} + m^{2})\Phi =0
\end{equation}
the action integral \eqref{actI1} is 
\begin{equation}
I[\Phi] = \frac{1}{2}\int_{z = \epsilon} (dx)_{d}\,\left(\frac{R}{z}\right)^{d-1}\Phi\partial_{z}\Phi
\end{equation}

The equation of motion \eqref{phiEq} is the elliptic equation, hence the solution of which is specified by the following choice of the boundary condition \cite{Witten:1998qj, Gubser:1998bc}
\begin{equation}
\label{bound}
\Phi(z, {\bf x}) \sim z^{d-\Delta}\Phi_{0}({\bf x})
\end{equation}
when $z=\epsilon\rightarrow 0^{+}$, $\Phi_{0}$ is an arbitrary function and  \cite{Witten:1998qj, Gubser:1998bc}
\begin{equation}
\label{bdelta}
\Delta = \frac{d}{2} \pm \sqrt{\frac{d^{2}}{4} + m^{2} R^{2}}
\end{equation}
Incidentally we notice here that equation \eqref{bdelta} can be obtained from the respective formula \eqref{Delta} for a massive scalar field in de Sitter space by replacing $R^{2}\rightarrow -R^{2}$  \cite{Antoniadis:1997fu,MaMot_PR_D2001,Casher:2003gc}.
The choice of boundary conditions \eqref{bound} leads to the following expression for the action functional \cite{Witten:1998qj, Gubser:1998bc}
\begin{equation}
\label{bterm}
I[\Phi_{0}] = \text{const}\; \int (dx)_{d}(dy)_{d}\; \frac{\Phi_{0}({\bf x})\Phi_{0}({\bf y})}{|{\bf x}-{\bf y}|^{2\Delta}}
\end{equation}
which indicates the direct relationship of the Euclidean QFT on $(d+1)$-dimensional Lobachevski space  with Euclidean $d$-dimensional CFT.  
In fact $e^{-I[\Phi_{0}]}$ is the classical approximation to the Feynman functional integral representation of the so-called partition function \cite{Witten:1998qj, Gubser:1998bc}
\begin{equation}
Z_{AdS}[\Phi_{0}] = \int_{\Phi_{0}}\mathcal{D}\Phi \exp{(-I[\Phi])}
\end{equation}
where the subscript on the functional integral indicates that one should integrate over field configurations $\Phi$ with the boundary condition \eqref{bound},
but at the same time it is the generating functional of the correlation functions of the quasi-primary operator $\mathcal{O}$ with scaling dimension $\Delta$ in some $\text{CFT}_{d}$  \cite{Witten:1998qj, Gubser:1998bc}
\begin{equation}
\label{adscft1}
Z_{CFT}[\Phi_{0}] = < \exp{\int (dx)_{d}\,\Phi_{0}\,\mathcal{O}} >
\end{equation}

The equation 
\begin{equation}
\label{adscft2}
Z_{AdS}[\Phi_{0}]  = Z_{CFT}[\Phi_{0}] 
\end{equation}
and the prescription described above is the basis of the conjecture known as the AdS/CFT correspondence \cite{Witten:1998qj, Gubser:1998bc}.
As we can see the only new element added to the previously known relationship between QFT on $(d+1)$-dimensional de Sitter space and CFT on $S^{d}$ resulting in the conjecture of the AdS/CFT correspondence is the introduction of the boundary terms in the action integral  \cite{Witten:1998qj,Gubser:1998bc}. The AdS/CFT correspondence prescription has been generalized to the case of the spinor, vector and and other tensor fields, for example in \cite{Mueck:1998iz,Henningson:1998cd}.

Over the past twelve years the AdS/CFT correspondence has been studied from various points of view leading to several significant results. 
Indeed, in the gravastar scenario for the physical black holes \cite{MaMotPNAS}, the black hole horizon of the exterior Schwarzschild solution and the cosmological horizon of the interior de Sitter space are replaced by the thin surface layer in which effectively massless degrees of freedom propagate.
Taking into account the facts that the interior of the gravastar is the de Sitter ball and the dS/CFT correspondence \cite{MaMot_PR_D2001} between QFT in de Sitter space and CFT on the boundary we should expect the emergence of the CFT behavior on the horizon. In the following sections we shall illustrate this observation by considering for simplicity massless fermions in the interior of the gravastar that is described by de Sitter ball.

In order to make the CFT near-horizon behavior of fermions propagating in the interior of the gravastar particularly transparent we shall introduce the concept of the optical geometry of de Sitter space.

\section{The optical geometry of de Sitter space}\label{sec:FermionsB}

Two metrics $g_{\mu\nu}$ and $g^{\prime}_{\mu\nu}$ belong to the same conformal equivalence class if they are related by the Weyl local conformal rescaling 
\begin{equation}
\label{weyl}
g^{\prime}_{\mu\nu}(x) = \Omega^{2}(x)\, g_{\mu\nu}(x)
\end{equation}
The local Weyl rescaling preserves null vectors $k^{\mu}$ because if $k^{\mu}g_{\mu\nu}k^{\nu} =0$  then $k^{\mu}g^{\prime}_{\mu\nu}k^{\nu} =0$ too.

For a massless particle the wave equation 
\begin{equation}
\label{WavEq}
\hat{W}(g)\Psi = 0 \; ,
\end{equation}
where $\hat{W}$ is the wave operator, is conformally invariant if 
\begin{equation}
\label{WavEqPr}
\hat{W}(g^{\prime})\Psi^{\prime} = \Omega^{-\Delta -2}\hat{W}(g)\Psi\; .
\end{equation}
with $\Delta$ the conformal weight of the massless integer spin field $\Psi$, that is $\Psi^{\prime} = \Omega^{-\Delta}\Psi$. For a half-integer spin 
\begin{equation}
\label{WavEqHa}
\hat{W}(g^{\prime})\Psi^{\prime} = \Omega^{-\Delta -1}\hat{W}(g)\Psi\; .
\end{equation}

The well known examples of conformally invariant wave equations for massless integer spin fields $\Psi$ are: 

the spin zero scalar field $\Phi$ with $\Delta=1$ and 
\begin{equation}
(-D^{\mu}D_{\mu} + \frac{R}{6})\Phi = 0
\end{equation}

and the Maxwell equations with $\Delta=0$ and 
\begin{equation}
-D^{2}A_{\mu} +D_{\mu}D^{\nu}A_{\nu} + R_{\mu}^{\nu}A_{\nu} =0
\end{equation}

while for a half-integer spin we have the Dirac equation for a massless spinor $\Psi$, for which $\Delta = 3/2$ and 
\begin{equation}
\imath \gamma^{\mu}D_{\mu}\Psi = 0 \, .
\end{equation}

It is well known that the transition from the wave optics to the geometric optics is accomplished by the eikonal approximation  
\begin{equation}
\Psi = A e^{\imath S}
\end{equation}
where $S$ is the eikonal. The hypersurfaces of constant $S$ are the wavefronts, while the integral lines $\frac{dx^{\mu}}{d\lambda} =k^{\mu} $ of the normal vector to the wavefronts $k^{\nu} = g^{\mu\nu}D_{\mu}S$ are called rays (light rays for the Maxwell equations). The conformally invariant wave equation  reduces in the eikonal approximation to the condition that $k^{\mu}$ is null and its integral lines are geodesics. Null lines are, of course, conformally invariant. Conformal invariance of the wave equation for massless particles (photons) is particularly useful because it allows to introduce a specific conformally related metric, known as an optical metric in which geometric optics can be derived from an analog of the Fermat principle. Later in this section we will describe an application of the optical geometry of de Sitter space to the massless Dirac field.

Consider  now a static spacetime with the metric 
\begin{equation}
\label{stat}
ds^{2} = g_{\mu\nu} dx^{\mu}dx^{\nu} = f \,dt^{2} - \gamma_{ij}dx^{i}dx^{j}
\end{equation}
where $f$ and $\gamma_{ij}$ are time independent.

A metric $g^{\prime}_{\mu\nu} = f^{-1} g_{\mu\nu}$ conformally related to \eqref{stat} is called the optical metric.
\begin{equation}
ds_{opt}^{2} = dt^{2} - h_{ij} dx^{i}dx^{j}
\end{equation}
where $h_{ij} = f^{-1}\gamma_{ij}$.

We will now present the optical metric of de Sitter space and analyze geometrical properties of the optical geometry of de Sitter space. 
De Sitter metric in the canonical static form is given by 
\begin{equation}
ds^{2} = (1-H^{2} r^{2}) dt^{2} - \frac{dr^{2}}{1-H^{2} r^{2}} -r^{2} (d\theta^{2}+\sin^{2}\theta d\varphi^{2})
\end{equation}

The optical metric for de Sitter space is now
\begin{equation}
ds_{opt}^{2} =  dt^{2} - \frac{dr^{2}}{(1-H^{2} r^{2})^{2}} -\frac{r^{2}}{1-H^{2} r^{2}} (d\theta^{2}+\sin^{2}\theta d\varphi^{2})
\end{equation}

The spatial part of the optical metric for de Sitter space
\begin{equation}
d\ell^{2} =   \frac{dr^{2}}{(1-H^{2} r^{2})^{2}} +\frac{r^{2}}{1-H^{2} r^{2}} (d\theta^{2}+\sin^{2}\theta d\varphi^{2})
\end{equation}
is conformally flat and it describes the geometry of 3-dimesional Lobachevski space with the constant negative curvature.

Indeed this can be seen by introducing coordinates $x = H r$, $y = H \rho$ and writing the spatial part of the optical metric in two conformally related forms
\begin{equation}
\label{spopt}
H^{2}d\ell^{2} = \frac{1}{(1-x^{2})^{2}} dx^{2} + \frac{x^{2}}{1-x^{2}}d\Omega^{2} =
\Phi^{2}(y) (dy^{2}+y^{2}d\Omega^{2})\qquad\qquad
\end{equation}
and using equations following from \eqref{spopt}
\begin{equation}
\label{eq1}
(1-x^{2})^{-1} dx = \Phi(y)dy
\end{equation}

\begin{equation}
\label{eq2}
x\,(1-x^{2})^{-1/2}  = y\,\Phi(y)
\end{equation}
which upon integration lead to the following result
\begin{equation}
\label{eq3}
x = \frac{2y}{1+y^{2}}
\end{equation}
and 
\begin{equation}
\label{eq4}
\Phi(y) = \frac{2}{1-y^{2}}
\end{equation}
where $0\leq x\leq 1$ and $0\leq y\leq 1$. The horizon at $r = H^{-1} \equiv R$ is mapped to $y=1$, which is  the boundary of the unit ball.

The spatial part of the optical metric
\begin{equation}
\label{poinc1}
d\ell^{2} = \frac{4 R^{2}}{(1-y^{2})^{2}} (dy^{2} + y^{2}d\Omega^{2}) = \frac{4 R^{2}d\textbf{y}\cdot d\textbf{y}}{(1-|\textbf{y}|^{2})^{2}}
\end{equation}
is described by the Poincare unit ball representation of the Lobachevski geometry, where $\textbf{y} =(y_{1},y_{2},y_{3})$.
Changing variables $y=\tanh(\psi/2)$ we obtain the usual form of the metric on the Lobachevski hyperbolic space $\textbf{H}^{3}$
\begin{equation}
d\ell^{2} = R^{2}(d\psi^{2} + \sinh^{2}\psi\,d\Omega^{2})
\end{equation}

This demonstrates that metric on de Sitter space is conformally related to the metric on the hyperbolic cylinder $\mathbb{R}\times \textbf{H}^{3}$. One can transform the Poincare metric on the unit ball \eqref{poinc1} to yet another Poincare form of the metric on the upper half-space $z\geqslant 0$
\begin{equation}
\label{poinc2}
d\ell^{2} = \frac{R^{2}}{z^{2}}(dx^{2}+dy^{2}+dz^{2})
\end{equation}
by the coordinate transformation
\begin{equation}
x = \frac{2y_{1}}{{y_{1}}^{2}+{y_{2}}^{2}+(1-y_{3})^{2}}
\end{equation}

\begin{equation}
y = \frac{2y_{2}}{{y_{1}}^{2}+{y_{2}}^{2}+(1-y_{3})^{2}}
\end{equation}

\begin{equation}
z = \frac{1-{y_{1}}^{2}-{y_{3}}^{2}-{y_{3}}^{2}}{{y_{1}}^{2}+{y_{2}}^{2}+(1-y_{3})^{2}}
\end{equation}

\section{The emergence of CFT behavior on de Sitter horizon}\label{sec:FermionsC}
As advertised above we now will describe an application of the optical geometry of de Sitter space to the massless Dirac field.  The concept of the optical geometry of de Sitter space is a particularly useful tool in uncovering the CFT behavior of matter fields on the cosmological horizon.  The essential point is that the optical geometry for the de Sitter space is the standard metric on the Cartesian product $\mathbb{R}\times H^{3}$, 
\begin{equation}
\label{opt}
ds_{opt}^{2} = dt^{2}-\frac{R^{2}}{z^{2}}(dx^{2}+dy^{2}+dz^{2})
\end{equation}
where $H^{3}$ is the 3-dimensional Euclidean anti-de Sitter (or Lobachevski) space. The cosmological horizon is at $z=0$ which is a 2-sphere. The appearance of the $AdS_{3}$ component in the optical geometry
of de Sitter space suggests that the slightly modified AdS/CFT prescription should be applicable to uncover the emergence of the CFT behavior of matter fields on the cosmological event horizon. 

The de Sitter metric is 
\begin{equation}
\label{opt2}
ds^{2} = \Omega^{2} ds_{opt}^{2}
\end{equation}
where the conformal factor $\Omega$ 
\begin{equation}
\label{conff}
\Omega =  \frac{2R\,z}{x^{2}+y^{2}+z^{2}+R^{2}}
\end{equation}
vanishes on the horizon $z=0$. The AdS/CFT correspondence for massive Dirac fermions was studied in \cite{Mueck:1998iz,Henningson:1998cd}

We now will consider the massless Dirac spinor field $\Psi$ on de Sitter space described by the action
\begin{equation}
\label{actio1}
I = \int_{\mathcal{M}} d^{4}x \sqrt{-g}\;\bar{\Psi}i\,\gamma^{\mu}D_{\mu}\Psi +\int_{\partial \mathcal{M}} d^{3}x \sqrt{h}\,\bar{\Psi}\,i\gamma_{3}\Psi\; .
\end{equation}

The covariant derivative $D_{\mu}$ acting on spinors is defined 
\begin{equation}
D_{\mu}\Psi = (\partial_{\mu}+ \omega_{\mu})\Psi = (\partial_{\mu}+\frac{1}{2}\omega^{ab}_{\mu}\Sigma_{ab})\Psi
\end{equation}
where $\omega^{ab}_{\mu}$ is the spin connection and $\Sigma^{ab} = \frac{1}{4}[\gamma^{a},\gamma^{b}]$.
The Dirac $\gamma$-matrices satisfy the standard anticommutation relations $\gamma^{a}\gamma^{b}+\gamma^{b}\gamma^{a} = 2\eta^{ab}$ where $\eta^{ab} = \text{diag}(+1,-1,-1,-1)$. One needs vierbein $e^{a}_{\mu}$ and its inverse $E^{\mu}_{a}$ in order to define the spin connection $\omega^{ab}_{\mu}$ and the position dependent Dirac matrices $\gamma^{\mu}$
\begin{equation}
\label{row1}
g_{\mu\nu} = \eta_{ab}e^{a}_{\mu}e^{b}_{\nu}
\end{equation}

\begin{equation}
\label{row2}
g^{\mu\nu} = \eta^{ab}E^{\mu}_{a}E^{\nu}_{b}
\end{equation}

\begin{equation}
\label{row3}
\gamma^{\mu} = E^{\mu}_{a}\gamma^{a}
\end{equation}

\begin{equation}
\label{row4}
D_{\mu}e^{a}_{\nu} = \partial_{\mu}e^{a}_{\nu} -\Gamma^{\lambda}_{\mu\nu}e^{a}_{\lambda} +\omega^{a}_{b\mu}e^{b}_{\nu} = 0
\end{equation}
Using the relations 

\begin{equation}
e^{a}_{\mu}E^{\mu}_{b} = \delta^{a}_{b}\qquad\qquad   E^{\mu}_{a} e^{a}_{\nu}= \delta^{\mu}_{\nu}
\end{equation}
applied to \eqref{row4} we get the formula
\begin{equation}
\omega^{a}_{b\mu} = \Gamma^{\lambda}_{\mu\nu}E^{\nu}_{b}e^{a}_{\lambda} - E_{b}^{\nu}\partial_{\mu}e^{a}_{\nu}
\end{equation}

The conformal invariance of massless Dirac fermions means that we can use the optical geometry for de Sitter space by substituting  $\Psi = \Omega^{-3/2} \Psi_{opt}$ and $g_{\mu\nu} = \Omega^{2} g_{\mu\nu}^{opt} $  in \eqref{actio1} obtaining 
\begin{equation}
\label{action2}
I = \int_{\mathcal{M}} d^{4}x \sqrt{-g_{opt}}\;\bar{\Psi}_{opt}i\,\gamma_{opt}^{\mu}D_{\mu}\Psi_{opt} +\int_{\partial \mathcal{M}} d^{3}x \sqrt{h_{opt}}\,\bar{\Psi}_{opt}\,i\gamma_{3}\Psi_{opt}\; .
\end{equation}

The boundary $\partial\mathcal{M}$ in the optical geometry is at $z = \epsilon\rightarrow 0^{+} $.
For the optical geometry \eqref{opt} we find the vierbein 
\begin{equation}
\label{vier}
e^{0}_{\mu} = \delta^{0}_{\mu} \qquad e^{m}_{\mu}= R\,z^{-1} \delta^{m}_{\mu}
\end{equation}
where $m=1, 2, 3$.
The spin connection $\omega_{\mu}$ is
\begin{equation}
\label{omeg}
\omega_{0} = \omega_{3} = 0\qquad\qquad \omega_{1} = z^{-1}\Sigma_{13}\qquad\qquad \omega_{2} = z^{-1}\Sigma_{23}
\end{equation}

In the following we will drop the subscript \textit{opt} in all formulae.

Using formulae \eqref{vier} and \eqref{omeg} we obtain the massless Dirac equation for an optical spinor $\Psi$ in the optical geometry
\begin{equation}
\label{Dir}
(\gamma^{0}\partial_{0}+R^{-1}z \gamma^{i}\partial_{i} - R^{-1}\gamma^{3})\Psi = 0
\end{equation}
The conjugate spinor $\bar{\Psi} = \Psi^{\dag}\gamma^{0}$ satisfies the following Dirac equation
\begin{equation}
\label{conDir}
\bar{\Psi}(\gamma^{0}\overleftarrow{\partial}_{0}+R^{-1}z \gamma^{i}\overleftarrow{\partial_{i}} - R^{-1}\gamma^{3}) = 0
\end{equation}
where we have used the chiral representation of Dirac gamma matrices for which 
\begin{equation}
{\gamma^{0}}^{\dag} = \gamma^{0}\qquad \qquad {\gamma^{i}}^{\dag} = - \gamma^{i}
\end{equation}

Rewriting the Dirac equation in the form 
\begin{equation}
\label{rewDir}
(z \gamma_{i}\partial_{i}-\gamma_{3})\Psi = R \gamma_{0}\partial_{0}\Psi
\end{equation}
where we have used the relations 
\begin{equation}
\gamma^{0}=\gamma_{0}\qquad\qquad \gamma^{i}= -\gamma_{i}
\end{equation}
we obtain the second order in derivatives of $\Psi$  equation 
\begin{equation}
\label{psieq}
(z^{2} \vec{\partial}^{2}+z^{2}\partial_{z}^{2} -2z \partial_{z} +2 +R^{2}\partial_{0}^{2} -R \gamma_{3}\gamma_{0}\partial_{0})\Psi =0
\end{equation}
where $\vec{\partial}= (\partial_{x}, \partial_{y})$ is the 2-diensional gradient operator.

We are looking for solutions to \eqref{psieq} in the form  \cite{Mueck:1998iz,Henningson:1998cd}
\begin{equation}
\label{psii}
\Psi(x) = \int \frac{d\omega d^{2}k}{(2\pi)^{3}} e^{-i\omega t + i {\bf k}\cdot {\bf x}}\Psi(\omega,{\bf k}; z)
\end{equation}
where $x = (t, x, y, z)$.

From equations \eqref{psieq} and \eqref{psii} we obtain an ordinary differential equation for the Fourier modes
$\Psi(\omega,{\bf k}; z)$
\begin{equation}
z^{2}\Psi^{\prime\prime}-2z \Psi^{\prime} +(2-\omega^{2}R^{2}+i\omega R \gamma_{3}\gamma_{0}-k^{2}z^{2})\Psi =0
\end{equation}
where $k = \lvert{\bf k}\rvert$. It is easy to see that the matrix $\gamma_{3}\gamma_{0}$ squares to $\mathbb{I}$, that is $(\gamma_{3}\gamma_{0})^{2}=\mathbb{I}$ and therefore its eigenvalues are $\pm1$.
Denoting the eigenspinors of $\gamma_{3}\gamma_{0}$ corresponding to eigenvalues $+1$ and $-1$, respectively as $\Psi_{+}$ and $\Psi_{-}$ we obtain
\begin{equation}
\label{eigP}
z^{2}\Psi_{\pm}^{\prime\prime}-2z \Psi_{\pm}^{\prime} +(2-\omega^{2}R^{2}\pm i\omega R -k^{2}z^{2})\Psi_{\pm} =0
\end{equation}

The following substitution
\begin{equation}
\Psi_{\pm} = z^{3/2} \Phi_{\pm}
\end{equation}
reduces \eqref{eigP} to the modified Bessel equation
\begin{equation}
\label{Bessel}
z^{2} \Phi_{\pm}^{\prime\prime} +z \Phi_{\pm}^{\prime} -(k^{2}z^{2}+\nu_{\pm}^{2})\Phi_{\pm} =0
\end{equation}
where $\nu_{\pm} = i\omega R \mp \frac{1}{2}$. Solutions to this Bessel equation \eqref{Bessel} are the modified Bessel functions known as the McDonald functions $K_{\nu_{\pm}}(kz)$ and $I_{\nu_{\pm}}(kz)$
\begin{equation}
\label{Ppm}
\Phi_{\pm} = a_{\pm}K_{\nu_{\pm}}(k\,z) + b_{\pm}I_{\nu_{\pm}}(k\,z) 
\end{equation}

It is useful to recall for future reference basic formulae for the McDonald functions
\begin{equation}
I_{\nu}(z) = (\frac{z}{2})^{\nu}\sum_{n=0}^{\infty}\frac{1}{n!}\frac{1}{\Gamma(\nu+n+1)}(\frac{z}{2})^{2n}
\end{equation}
\begin{equation}
K_{\nu}(z) = \frac{\pi}{2\sin(\pi\nu)}(I_{-\nu}(z)-I_{\nu}(z)) \qquad\qquad K_{-\nu} = K_{\nu}
\end{equation}
For small $z\rightarrow 0$ the asymptotics of $I_{\nu}(z)$ and $K_{\nu}(z)$ are
\begin{align}
I_{\nu}(z)\sim &\frac{1}{\Gamma(\nu+1)}\left(\frac{z}{2} \right)^{\nu}\\
\label{Kasymp} K_{\nu}(z)\sim  &\frac{\pi}{2\sin(\pi\nu)}\left[\frac{1}{\Gamma(-\nu+1)}\left(\frac{z}{2} \right)^{-\nu} - \frac{1}{\Gamma(\nu+1)}\left(\frac{z}{2} \right)^{\nu} \right]
\end{align}
Using the well known relation for the Euler gamma function
\begin{equation}
\Gamma(\nu)\Gamma(1-\nu)=\frac{\pi}{\sin(\pi\nu)}
\end{equation}
we can write \eqref{Kasymp} in the form
\begin{equation}
\label{Kasymp2}
K_{\nu}(z)\sim  \frac{1}{2}\left[\Gamma(\nu)\left(\frac{z}{2} \right)^{-\nu} + \Gamma(-\nu)\left(\frac{z}{2} \right)^{\nu} \right]
\end{equation}

Physically acceptable solutions are those that are regular when $z \rightarrow +\infty$. It is because $I_{\nu}(z)$ behaves as $e^{z}$ while $K_{\nu}(z)$ behaves as $e^{-z}$ when $z\rightarrow +\infty$ we have to set $b_{\pm} =0$ in \eqref{Ppm}. Thus the solution for the Fourier modes is
\begin{equation}
\Psi_{\pm}(\omega,{\bf k};z) = z^{3/2} a_{\pm} K_{\nu_{\pm}}(k\,z)
\end{equation}
where 
\begin{equation}
\nu_{\pm} = i\omega R \mp \frac{1}{2}
\end{equation}

The eigenspinors $a_{\pm}$ of the matrix $\gamma_{3}\gamma_{0}$ depend on ${\bf k}$ vector and 
satisfy the relation 
\begin{equation}
\label{gammarel}
\gamma_{0}\;a_{\pm} = \mp \gamma_{3}\;a_{\pm}
\end{equation}
Substituting the solution
\begin{equation}
\Psi =z^{3/2}\Phi = z^{3/2}e^{-i\omega t + i\,{\bf k}\cdot{\bf x}}(a_{+}K_{\nu_{+}}(kz)+a_{-}K_{\nu_{-}}(kz))
\end{equation}
to the Dirac equation \eqref{rewDir} we obtain
\begin{equation}
\label{Dirag}
\left[iz\,\boldsymbol{\gamma}\cdot{\bf k} + \gamma_{3}(z\partial_{z}+\frac{1}{2})\right]\Phi = -i\,\omega R \gamma_{0}\Phi
\end{equation}
Writing 
\begin{equation}
\Phi = \Phi_{+} +\Phi_{-}
\end{equation}
where
\begin{equation}
\Phi_{+} =e^{-i\omega t + i\,{\bf k}\cdot{\bf x}}a_{+}K_{\nu_{+}}(kz)
\end{equation}
\begin{equation}
\Phi_{-} =e^{-i\omega t + i\,{\bf k}\cdot{\bf x}}a_{-}K_{\nu_{-}}(kz)
\end{equation}
and using the relation \eqref{gammarel} the equation \eqref{Dirag} assumes the form
\begin{equation}
\label{Dirag2}
iz\,\boldsymbol{\gamma}\cdot{\bf k}(\Phi_{+}+\Phi_{-})+\gamma_{3}(z\partial_{z}-\nu_{+})\Phi_{+}+\gamma_{3}(z\partial_{z}+\nu_{-})\Phi_{-} =0
\end{equation}
where $\boldsymbol{\gamma} = (\gamma_{1}, \gamma_{2})$.

Using the following relations for the McDonald's function $K_{\nu}(z)$
\begin{equation}
K_{\nu-1} - K_{\nu+1} = -\frac{2\nu}{z}K_{\nu}
\end{equation}
\begin{equation}
K_{\nu-1} + K_{\nu+1} = -2\frac{d}{dz}K_{\nu}
\end{equation}
we obtain
\begin{equation}
\label{zdz1}
(z \frac{d}{dz} + \nu)K_{\nu} = -z K_{\nu-1}
\end{equation}
\begin{equation}
\label{zdz2}
(z \frac{d}{dz} - \nu)K_{\nu} = -z K_{\nu+1}
\end{equation}
We now notice that
\begin{equation}
\label{rela1}
\nu_{-}-1 = \nu_{+}
\end{equation}

Applying relations \eqref{zdz1} --- \eqref{rela1} to \eqref{Dirag2} we now obtain the following equation
\begin{equation}
iz\,\boldsymbol{\gamma}\cdot{\bf k} (a_{+}K_{\nu_{+}}+a_{-}K_{\nu_{-}})-k\,z\gamma_{3}(a_{+}K_{\nu_{-}}+a_{-}K_{\nu_{+}}) = 0
\end{equation}
that leads to the direct relation between eigenspinors $a_{+}$ and $a_{-}$
\begin{equation}
a_{-} = \frac{i\,\boldsymbol{\gamma}\cdot{\bf k}}{k}\gamma_{3}\,a_{+}
\end{equation}
\begin{equation}
a_{+} = \frac{i\,\boldsymbol{\gamma}\cdot{\bf k}}{k}\gamma_{3}\,a_{-}
\end{equation}
In order to demonstrate the correspondence between the Quantum Field Theory of the massless Dirac field in de Sitter space and the Conformal Field Theory on de Sitter cosmological  horizon, which may be called the dS/CFT correspondence, we follow by analogy the AdS/CFT prescription  \cite{Mueck:1998iz,Henningson:1998cd} given in \eqref{adscft1} and \eqref{adscft2}. 
We expect that the Lorentzian partition function for massless fermions evaluated on de Sitter space will depend on the boundary spinors $\Psi_{0}$, $\bar{\Psi}_{0}$ defined properly on the de Sitter horizon
\begin{equation}
\label{zds}
Z_{dS}[\Psi_{0}, \bar{\Psi}_{0}] = \int_{\Psi_{0}, \bar{\Psi}_{0}}\mathcal{D}\Psi\mathcal{D}\bar{\Psi}\;\exp{(i I[\Psi,\bar{\Psi}])}
\end{equation}
where the subscript on the functional integral indicates that one should integrate over field 
cofigurations $\Psi,\bar{\Psi}$ with $\Psi_{0}$ and $\bar{\Psi}_{0}$ as their boundary values on the de Sitter horizon.

In analogy to the AdS/CFT prescription we put forward the following conjecture 
\begin{equation}
\label{zds2}
Z_{dS}[\Psi_{0}, \bar{\Psi}_{0}] = Z_{CFT}[\Psi_{0}, \bar{\Psi}_{0}] 
\end{equation}

\begin{equation}
Z_{CFT}[\Psi_{0}, \bar{\Psi}_{0}] = < \exp [i \int (dx)_{3} (\bar{\Psi}_{0}\,\mathcal{O}+\bar{\mathcal{O}}\Psi_{0})] >
\end{equation}
where $\mathcal{O}$ and $\bar{\mathcal{O}}$ are quasi-primary spinor operators in CFT on de Sitter horizon, and $(dx)_{3} = dt d^{2}x$.

In order to implement the dS/CFT prescription we now need to evaluate $Z_{dS}[\Psi_{0}, \bar{\Psi}_{0}]$ given by the Gaussian functional integral \eqref{zds}. This Gaussian functional integral can be computed explicitly in terms of the boundary term in the action \eqref{actio1} evaluated on the solutions to the Dirac equation for   $\Psi$ and $\bar{\Psi}$ that depend functionally on the boundary spinors $\Psi_{0}, \bar{\Psi}_{0}$.

For solutions of the Dirac equation \eqref{Dir} the action \eqref{action2} evaluates to the boundary term
\begin{equation}
\label{DiracAct}
I[\Psi] = \int_{z=\epsilon} dt d^{2}x R^{2} z^{-2} \bar{\Psi}(t,\vec{x},z)\,i\gamma_{3}\Psi(t,\vec{x},z) 
\end{equation}

The solution to the Dirac equation can now be expressed in terms of the boundary spinor mode $\Psi_{-,\epsilon}\equiv \Psi_{-}(\omega,{\bf k}; z=\epsilon)$ by using the relations
\begin{equation}
a_{+} = \epsilon^{-3/2}  \frac{i\,\boldsymbol{\gamma}\cdot{\bf k}}{k}\gamma_{3}\frac{\Psi_{-,\epsilon}(\omega,{\bf k})}{K_{\nu_{-}}(k\epsilon)}
\end{equation}
\begin{equation}
a_{-} = \epsilon^{-3/2} \frac{\Psi_{-,\epsilon}(\omega,{\bf k})}{K_{\nu_{-}}(k\epsilon)}
\end{equation}

\begin{equation}
\label{PsiB}
\Psi(x) = \int \frac{d\omega d^{2}k}{(2\pi)^{3}} e^{-i\omega t + i {\bf k}\cdot {\bf x}}\left(\frac{z}{\epsilon}\right)^{3/2}\left[\frac{i\,\boldsymbol{\gamma}\cdot{\bf k}}{k}\gamma_{3}K_{\nu_{+}}(kz) + K_{\nu_{-}}(kz) \right]\frac{\Psi_{-,\epsilon}(\omega,{\bf k})}{K_{\nu_{-}}(k\epsilon)}
\end{equation}

The solution to the Dirac equation for the conjugate spinor $\bar{\Psi}(x)$ can be similarly expressed in terms of the boundary spinor mode $\bar{\Psi}_{+,\epsilon}\equiv \bar{\Psi}_{+}(\omega,{\bf k}; z=\epsilon)$

\begin{equation}
\bar{\Psi}_{+}(\omega,{\bf k}; z=\epsilon) = \left(\Psi_{-,\epsilon}(-\omega,- {\bf k})\right)^{\dag} \gamma^{0}
\end{equation}

\begin{equation}
\label{PsibB}
\bar{\Psi}(x) = \int \frac{d\omega d^{2}k}{(2\pi)^{3}} e^{-i\omega t + i {\bf k}\cdot {\bf x}}\left(\frac{z}{\epsilon}\right)^{3/2}\frac{\bar{\Psi}_{+,\epsilon}(\omega,{\bf k})}{K_{\nu_{-}}(k\epsilon)}\left[K_{\nu_{-}}(kz)+\gamma_{3}\frac{i\,\boldsymbol{\gamma}\cdot{\bf k}}{k}K_{\nu_{+}}(kz)  \right]
\end{equation}

We compute the boundary term \eqref{DiracAct} in the momentum space by substituting expressions \eqref{PsiB} and \eqref{PsibB} into it
\begin{equation}
\label{boundmom}
I = i (2\pi)^{-3}(\frac{R}{\epsilon})^{2}\int d\omega d^{2}k \bar{\Psi}_{+, \epsilon}(\omega, {\bf k})\frac{i\boldsymbol{\gamma}\cdot{\bf k}}{k} \Psi_{-, \epsilon}(-\omega, -{\bf k})\left[\frac{K_{\nu_{+}}(k\epsilon)}{K_{\nu_{-}}(k\epsilon)}-\frac{K_{\nu_{-}}(k\epsilon)}{K_{\nu_{+}}(k\epsilon)}\right]
\end{equation}

Equivalent expression for the boundary term \eqref{boundmom} in the position space is (see, e.g.  \cite{Mueck:1998iz,Henningson:1998cd})
\begin{equation}
\label{boundpos}
I = i \int (dx)_{3}(dx^{\prime})_{3} \bar{\Psi}_{+, \epsilon}(x) S_{\epsilon}(x, x^{\prime})\Psi_{-, \epsilon}(x^{\prime})
\end{equation}
where the kernel $S_{\epsilon}$ in the integral \eqref{boundpos} is defined by the formulae
\begin{equation}
\label{kernel}
S_{\epsilon}(x, x^{\prime}) = (\frac{R}{\epsilon})^{2} \int \frac{d\omega d^{2}k}{(2\pi)^{3}} e^{i\omega (t-t^{\prime})-i{\bf k}\cdot({\bf x}-{}\bf x^{\prime})}\frac{i\boldsymbol{\gamma}\cdot{\bf k}}{k}\left[\frac{K_{\nu_{+}}(k\epsilon)}{K_{\nu_{-}}(k\epsilon)}-\frac{K_{\nu_{-}}(k\epsilon)}{K_{\nu_{+}}(k\epsilon)}\right]
\end{equation}
\begin{equation}
\label{kernel2}
S_{\epsilon}(x, x^{\prime}) = -(\boldsymbol{\gamma}\cdot\boldsymbol{\partial}) G_{\epsilon}(x, x^{\prime})
\end{equation}
where
\begin{equation}
\label{kernel3}
G_{\epsilon}(x, x^{\prime}) = (\frac{R}{\epsilon})^{2} \int \frac{d\omega d^{2}k}{(2\pi)^{3}} e^{i\omega (t-t^{\prime})-i{\bf k}\cdot({\bf x}-{}\bf x^{\prime})}\frac{1}{k}\left[\frac{K_{\nu_{+}}(k\epsilon)}{K_{\nu_{-}}(k\epsilon)}-\frac{K_{\nu_{-}}(k\epsilon)}{K_{\nu_{+}}(k\epsilon)}\right]
\end{equation}

In order to compute the limit $\epsilon \rightarrow 0^{+}$ in \eqref{boundpos} we now must evaluate the asymptotic behavior with $\epsilon$ in \eqref{kernel}. To this end we compute
\begin{equation}
\label{ratio1}
\frac{K_{\nu_{+}}(k\epsilon)}{K_{\nu_{-}}(k\epsilon)} \sim \frac{k\epsilon}{2} \frac{\Gamma(i\omega R -1/2)}{\Gamma(i\omega R +1/2)} + (\frac{k\epsilon}{2})^{2i\omega R}\frac{\Gamma(1/2-i\omega R)}{\Gamma(1/2+i\omega R)}
\end{equation}
and 
\begin{equation}
\label{ratio2}
\frac{K_{\nu_{-}}(k\epsilon)}{K_{\nu_{+}}(k\epsilon)} \sim \frac{k\epsilon}{2} \frac{\Gamma(-i\omega R -1/2)}{\Gamma(-i\omega R +1/2)} + (\frac{k\epsilon}{2})^{-2i\omega R}\frac{\Gamma(1/2+i\omega R)}{\Gamma(1/2-i\omega R)}
\end{equation}

The first term in \eqref{ratio1} and \eqref{ratio2} becomes $k$-independent when multiplied by the factor $1/k$ in the formula \eqref{kernel3}. This means that these terms contribute to the Dirac delta $\delta^{2}({\bf x}-{\bf x}^{\prime})$ in the kernel \eqref{kernel3}. These kind of terms in the Green's function are known as contact terms and we shall neglect them from now on. 

The second term in \eqref{ratio1} contributes a factor $\epsilon^{-2+2i\omega R}$ in \eqref{boundmom} that can be absorbed in the definition of the boundary spinors $\Psi_{0}$ and $\bar{\Psi}_{0}$
\begin{equation}
\label{bspindef}
\lim_{\epsilon\rightarrow 0^{+}}\epsilon^{-2+2i\omega R} \bar{\Psi}_{+, \epsilon}(\omega, {\bf k}) \Psi_{-, \epsilon}(-\omega, -{\bf k}) = \bar{\Psi}_{0}(\omega, {\bf k}) \Psi_{0}(-\omega, -{\bf k})
\end{equation}

After all powers of $\epsilon$ have been absorbed in the definition of the boundary spinors $\Psi_{0}$ and $\bar{\Psi}_{0}$ the action \eqref{boundpos} takes the form
\begin{equation}
\label{surfterm}
I = i \int (dx)_{3}(dx^{\prime})_{3} \bar{\Psi}_{0}(x) S(x, x^{\prime})\Psi_{0}(x^{\prime})
\end{equation}
where
\begin{equation}
S(x, x^{\prime}) = -(\boldsymbol{\gamma}\cdot\boldsymbol{\partial}) G(x, x^{\prime})
\end{equation}

The  kernel $G(x, x^{\prime})$ receives two contributions from \eqref{ratio1} and \eqref{ratio2}
\begin{equation}
G(x, x^{\prime}) = G_{+}(x, x^{\prime})+ G_{-}(x, x^{\prime})
\end{equation}
where $G_{+}(x, x^{\prime})$ is 
\begin{equation}
\label{Gplus}
G_{+}(x, x^{\prime}) =\frac{1}{2} R^{2} \int \frac{d\omega d^{2}k}{(2\pi)^{3}} e^{i\omega (t-t^{\prime})-i{\bf k}\cdot({\bf x}-{}\bf x^{\prime})}(\frac{k}{2})^{2i\omega R-1}\frac{\Gamma(1/2-i\omega R)}{\Gamma(1/2+i\omega R)}
\end{equation}
and 
\begin{equation}
\label{Gminus}
G_{-}(x, x^{\prime}) =-\frac{1}{2} R^{2} \int \frac{d\omega d^{2}k}{(2\pi)^{3}} e^{i\omega (t-t^{\prime})-i{\bf k}\cdot({\bf x}-{}\bf x^{\prime})}(\frac{k}{2})^{-2i\omega R-1}\frac{\Gamma(1/2+i\omega R)}{\Gamma(1/2-i\omega R)}
\end{equation}

In the definition of $G_{+}(x, x^{\prime})$ the usual $i\varepsilon$-prescription is understood: $\omega \rightarrow \omega -i\varepsilon$ where $\varepsilon \rightarrow 0^{+}$, while for $G_{-}(x, x^{\prime})$ we use $\omega \rightarrow \omega +i\varepsilon$. This is as it should be because in QFT the Green's functions are defined by the vacuum expectation value of the time-ordered chronological product of operators. For $G_{+}(x, x^{\prime})$ it produces the $\Theta (t^{\prime}-t)$ factor and for $G_{-}(x, x^{\prime})$ the $\Theta (t-t^{\prime})$ factor.

In order to evaluate the kernel $G(x, x^{\prime})$ we need the following Fourier transform  \cite{Mueck:1998iz,Henningson:1998cd}
\begin{equation}
\label{fourier}
\int (dk)_{d} (2\pi)^{-d} k^{2m-1} e^{-i{\bf k}\cdot{\bf x}} = 2^{2m-1} \pi^{-d/2} \frac{\Gamma(\frac{d-1}{2}+m)} {\Gamma(\frac{1}{2}-m)} r^{-(2m+d-1)}
\end{equation}
where $r = \lvert {\bf x}\rvert$. Applying \eqref{fourier} to \eqref{Gplus} in the case when $d=2$, and $m=i\omega R$ we obtain the following formula for $G_{+}(x,x^{\prime})$
\begin{equation}
G_{+}(x, x^{\prime}) =\left(\frac{R}{2\pi}\right)^{2} \int d\omega e^{i\omega (t-t^{\prime})}\lvert{\bf x}-{\bf x}^{\prime} \rvert^{-(2i\omega R +1)}
\end{equation}

In the similar fashion we obtain the formula for $G_{-}(x,x^{\prime})$
\begin{equation}
G_{-}(x, x^{\prime}) = - \left(\frac{R}{2\pi}\right)^{2} \int d\omega e^{-i\omega (t-t^{\prime})}\lvert{\bf x}-{\bf x}^{\prime} \rvert^{-(2i\omega R +1)}
\end{equation}

The dS/CFT correspondence prescription discussed above allows us to read off from \eqref{surfterm} the correlation function of the quasi primary operators $\mathcal{O}(x)$ and $\bar{\mathcal{O}}(x^{\prime})$
\begin{equation}
\label{finres}
< T(\mathcal{O}(x) \bar{\mathcal{O}}(x^{\prime})) > = \left(i \boldsymbol{\gamma}\cdot\boldsymbol{\partial}\right)G(x,x^{\prime})
\end{equation}
where the chronological product of fermionic operators is defined
\begin{equation}
\label{chrondef}
T(\mathcal{O}(x) \bar{\mathcal{O}}(x^{\prime})) = \Theta (t-t^{\prime})\mathcal{O}(x) \bar{\mathcal{O}}(x^{\prime}) - \Theta (t^{\prime}-t) \bar{\mathcal{O}}(x^{\prime})\mathcal{O}(x) 
\end{equation}

It should be noticed that two terms in the vacuum average of the chronological product \eqref{chrondef} are reproduced on the right hand side of the formula \eqref{finres} by the contributions coming from $G_{+}$ and $G_{-}$. This verifies that dS/CFT correspondence prescription conjectured above is basically correct. 
We have illustrated the dS/CFT correspondence on the example of the massless Dirac field. 
Our computation verifying the dS/CFT correspondence for massless Dirac fermions can be relatively easily generalized to the higher (and lower) spins. The dS/CFT correspondence for the cases of the scalar, vector, 
and tensor fields will be treated in the future publication(s).

\end{document}